# Triangle Distribution and Equation of State for Classical Rigid Disks

Dorothea K. Stillinger,[1]  Frank H. Stillinger,[1,2]  Salvatore Torquato,[2,3,4]

Thomas M. Truskett,[5] and Pablo G. Debenedetti[5]

[1]  Bell Laboratories, Lucent Technologies, Murray Hill, NJ 07974.

[2]  Princeton Materials Institute, Princeton University, Princeton, NJ 08544.

[3]  Department of Civil Engineering and Operations Research, Princeton University, Princeton, NJ, 08544.

[4]  Institute for Advanced Study, Princeton, NJ 08540.

[5]  Department of Chemical Engineering, Princeton University, Princeton, NJ 08544.

Abstract

The triangle distribution function $f^{(3)}$ for three mutual near neighbors in the plane describes basic aspects of short-range order and statistical thermodynamics in two-dimensional many-particle systems. This paper examines prospects for constructing a self-consistent calculation for the rigid-disk-system $f^{(3)}$. We present several identities obeyed by $f^{(3)}$. A rudimentary closure suggested by scaled-particle theory is introduced. In conjunction with three of the basic identities, this closure leads to an unique $f^{(3)}$ over the entire density range. The pressure equation of state exhibits qualitatively correct behaviors in both the low density and the close-packed limits, but no intervening phase transition appears. We discuss extensions to improved disk closures, and to the three-dimensional rigid sphere system.

I.  Introduction

Representing the statistical behavior of material systems with rigid particle models has a long and venerable history.[1-5]  Conceptually, the simplest of these models involves rigid spheres, or analogously rigid disks in two dimensions.  Modern interest focuses on the fluid-crystal phase transition,[6-8] and on the existence and properties of amorphous dense packings of these idealized "molecules".[9-13]

The present paper revisits the statistical geometry and thermodynamics of the single-component rigid disk system.  Virial coefficients are known through eighth order for this model,[14] and a long series of computer simulations (both Monte Carlo [7,8,15-17] and molecular dynamics [6,18,19]) have been devoted to determining its equation of state.  Nevertheless, some basic issues remain unresolved, including the thermodynamic order of the melting/freezing transition, and the related question about whether the model supports a stable "hexatic" phase.[20-24]

The strategy chosen here examines the statistics of near-neighbor particle triangles; these are generated by the Voronoi-Delaunay tessellation as explained in the following Section II.  Collins pioneered this approach for two-dimensional systems some years ago,[25,26] and we have managed to extend his work in several respects.  The definition of the basic triangle distribution function generated by the tessellation, and its elementary properties, form the content of Section III.  Some further useful identities obeyed by this distribution function appear in Section IV.  Section V examines the approach to the crystalline close-packed limit.

In an effort to build a closed-form predictive theory for the rigid disk equation of state, Section VI introduces and motivates an approximation for the triangle distribution function.

When used in connection with identities shown earlier, Sections III and IV, this generates a pair of nonlinear simultaneous equations in two scalar unknowns. Section VII presents the numerical procedure that has been applied to find solutions, and reveals its results. A final Section VIII discusses alternative closures, and considers the requirements for extension to rigid spheres in three dimensions, as well as to mixtures of particles of different sizes. An Appendix collects some explicit results for the triangle distribution function and its implied near-neighbor pair distribution function for the two-dimensional classical ideal gas.

## II. Voronoi-Delaunay Tessellation

Suppose $N$ particles (for the moment not necessarily rigid disks) reside at locations $\mathbf{r}_1$ … $\mathbf{r}_N$ within rectangular area $A$ in the plane. Furthermore, suppose that periodic boundary conditions apply, so that $A$ is surrounded by images of itself stretching to infinity in all directions. Near-neighbor (Voronoi) polygons can be constructed about each of the $N$ points and their images;[27,28] these are defined to contain all locations that are closer to the embedded point, at $\mathbf{r}_i$ say, than to any other $\mathbf{r}_j$ ($j \neq i$). As Figure 1 indicates by solid lines, the boundaries of the Voronoi polygons are composed of segments of the perpendicular bisectors of lines that connect the embedded point to neighboring points.

Any particle or periodic image $j$ whose bisector with $i$ contributes a boundary line segment to the Voronoi polygon surrounding $i$ is defined to be a near neighbor to $i$. Near-neighbor pairs are connected by dashed lines in Figure 1. The Delaunay tessellation, or tiling, of the plane consists of the triangles that are composed of near-neighbor links.[25-28] These triangles cover the plane (periodically, with the chosen boundary conditions) without gaps or overlaps.

The analogous three-dimensional version surrounds each given point with a near-neighbor (Voronoi) polyhedron whose boundary consists of polygonal sections of bisector planes. Each such polygonal section identifies a near neighbor for that point. The collection of near-neighbor pair links defines tetrahedra that tessellate, or tile, the three-dimensional space without gaps or overlaps.

Returning to two dimensions, the circumscribed circle for a triangle of mutual near neighbors $i, j, k$ has a special significance that is illustrated in Figure 2. In order that these three particles indeed be near neighbors to one another, no fourth particle can be permitted to enter the interior of that circumscribed circle.[25] The exterior position denoted "$l$" in Figure 2 is permissible, that interior position denoted "$l'$" is not, since it would eliminate the near-neighbor relation between $i$ and $j$. Thus the circumscribed circle can be viewed as an impenetrable barrier for all other particles. When it is expressed as a function of the triangle side lengths, the radius of the circle is

$$R = \frac{r_{ij} r_{ik} r_{jk}}{\left[2(r_{ij}^2 r_{ik}^2 + r_{ij}^2 r_{jk}^2 + r_{ik}^2 r_{jk}^2) - r_{ij}^4 - r_{ik}^4 - r_{jk}^4\right]^{1/2}} \quad . \tag{2.1}$$

The corresponding three-dimensional case involves the circumscribed sphere for each tetrahedron of four mutual near neighbors. No fifth particle position can be permitted to penetrate this sphere, for if it did at least one of the tetrahedron's six near-neighbor pairings would be violated. The circumscribed sphere acts as an absolute exclusion zone for its tetrahedron. In principle, the radius of the tetrahedron's circumscribed sphere can be expressed in closed form as a function of the six tetrahedron edge lengths. But unlike the two-dimensional R shown in Eq. (2.1), it cannot be a fully symmetric function of its variables, because the order in which those edges connect is crucial.

Returning to the specific case of rigid disks in the plane, no near-neighbor pair can have separation less than the collision diameter $a$. This model's rigid pair interaction also augments the exclusion circle for external particles shown in Figure 2 with impenetrable circular caps of radius $a$ centered at each of $i, j$, and $k$. When these mutual near neighbors are at or near rigid disk contact, only the circular caps are exposed; the circumscribed circle then is "hidden" from exterior particles $l$.

III. Triangle Distribution Function

Let $d\mathbf{r}_1$, $d\mathbf{r}_2$, and $d\mathbf{r}_3$ be three distinct area differentials in the plane. The probability that all three are simultaneously occupied by particles that are mutual near neighbors to one another will be denoted by

$$\boldsymbol{r}^3 f^{(3)}(\mathbf{r}_1,\mathbf{r}_2,\mathbf{r}_3)d\mathbf{r}_1 d\mathbf{r}_2 d\mathbf{r}_3 \quad . \tag{3.1}$$

Here the number density $N/A$ has been denoted by $\boldsymbol{r}$. The use of periodic boundary conditions automatically causes $f^{(3)}$ to possess translation invariance. In the infinite system limit that will be of interest in the following ($A$ diverging, while its shape and the number density $\boldsymbol{r}$ remain fixed), we know that $f^{(3)}$ will also be rotation invariant if the system remains in the fluid phase. It is not known at present whether the same rotation invariance obtains for the infinite-system-limit $f^{(3)}$ when the system is in the crystalline phase. However we are at liberty to interpret $f^{(3)}$ as an orientation-averaged probability function for the near-neighbor triangles, so regardless of phase it then becomes a function of the three triangle side lengths, $f^{(3)}(r_{12},r_{13},r_{23})$.

The function $f^{(3)}$ is analogous to the conventional three-particle correlation function $g^{(3)}$. The latter is defined by the statement, similar to Eq. (3.1) above, that

$$r^3 g^{(3)}(\mathbf{r}_1, \mathbf{r}_2, \mathbf{r}_3) d\mathbf{r}_1 d\mathbf{r}_2 d\mathbf{r}_3 \qquad (3.2)$$

is the probability for differential area elements $d\mathbf{r}_1$, $d\mathbf{r}_2$, and $d\mathbf{r}_3$ to be occupied simultaneously by any three particles.[29] Because definition Eq. (3.2) does not require a mutual near-neighbor relationship, $f^{(3)}$ and $g^{(3)}$ obviously must satisfy the inequalities

$$0 \leq f^{(3)} \leq g^{(3)} \ . \qquad (3.3)$$

for any particle-triplet configuration.

The distance variables $r_{12}$, $r_{13}$, and $r_{23}$ to be used for $f^{(3)}$ can be regarded as independent only if they satisfy the triangle inequalities, *i.e.* if they are geometrically capable of forming a triangle. One easily verifies that this condition is equivalent to positivity of the quartic multinomial appearing in the denominator of Eq. (2.1):

$$2(r_{12}^2 r_{13}^2 + r_{12}^2 r_{23}^2 + r_{13}^2 r_{23}^2) - r_{12}^4 - r_{13}^4 - r_{23}^4 > 0 \ . \qquad (3.4)$$

Whenever the three distances conspire to produce a zero-area triangle (the sum of two sides equals the third side), this multinomial vanishes and the radius of the circumscribed circle becomes infinite.

IV. Some Identities

The sum of internal angles for any triangle is $\pi$, while the sum of such vertex angles at each of the $N$ particles in the Voronoi-Delaunay tessellation is $2\pi$. Consequently, the number of triangles of mutual near neighbors must be exactly $2N$. This fact imposes the following normalization condition on the distribution function $f^{(3)}$:

$$2N = (\pi^3 / 3!) \int d\mathbf{r}_1 \int d\mathbf{r}_2 \int d\mathbf{r}_3\, f^{(3)} \quad . \tag{4.1}$$

Here the integrals each span the system area, and the denominator factor 3! compensates for the multiple counting of each triangle as $\mathbf{r}_1$, $\mathbf{r}_2$, and $\mathbf{r}_3$ independently sweep over that area. Equation (4.1) adopts a more useful form when translation and rotation invariance of $f^{(3)}$ are invoked, and triangle side lengths ( $r$, $s$, and $t$, hereafter) are used as a nonorthogonal set of integration variables. After including the proper transformation Jacobian, Eq. (4.1) leads to:

$$3/(2\pi r^2) = \int_a^\infty dr \int_a^\infty ds \int_a^\infty dt\, rst\, f^{(3)}(r,s,t) \left[ M(r,s,t) \right]^{-1/2} \quad , \tag{4.2}$$

where, as before, $a$ is the disk diameter, and $M$ is the quartic multinomial appearing in $R$ [Eq. (2.1)] and Eq. (3.4):

$$M(r,s,t) = 2(r^2 s^2 + r^2 t^2 + s^2 t^2) - r^4 - s^4 - t^4 \quad . \tag{4.3}$$

Furthermore, no triangles are possible unless inequality (3.4) is satisfied, so here and in the following, the product $f^{(3)} M^{-1/2}$ is interpreted as zero if $M \leq 0$.

We note in passing that normalization condition (4.2) is logically equivalent to the well-known result (often presented as an application of Euler's theorem [30]) that the average number of near neighbors in the plane is exactly 6. The total number of sides possessed by $2N$ triangles is $6N$. In the Voronoi-Delaunay tessellation each pair link serves as a shared side, so $3N$ pair links exist. But each link has 2 ends. Thus the total near neighbor count for the $N$ particles is $6N$.

Normalization condition (4.2) can be classified as a purely topological statement. We also require that a mensuration condition be satisfied, namely that the sum of the areas $\Delta(r,s,t)$ of all $2N$ triangles equal the system area $A$. One has

$$\Delta(r,s,t) = (1/4)\left[M(r,s,t)\right]^{1/2} \; ; \tag{4.4}$$

consequently the area condition reduces to:

$$3/(\boldsymbol{p}\boldsymbol{r}^3) = \int_a^\infty dr \int_a^\infty ds \int_a^\infty dt \; rst \; f^{(3)}(r,s,t) \; . \tag{4.5}$$

Figure 3 presents an extremal configuration of three disks that necessarily must be mutual near neighbors because their radius-$a$ exclusion zones entirely obscure the triangle's circumscribed circle. In this arrangement, with $r$, $s$, and $t$ all equal to $3^{1/2}a$ (and $R = a$), infinite disk repulsions prevent violation of the three near-neighbor pairings shown. This automatic nonviolation situation continues to exist for some set **U** of triples $r$, $s$, and $t$, including specifically all cases for which

$$a \leq r = s = t < 3^{1/2}a \; . \tag{4.6}$$

Therefore we have

$$f^{(3)}(r,s,t) = g^{(3)}(r,s,t) , \qquad (r,s,t) \in \mathbf{U} , \qquad (4.7)$$

a result noted earlier by Collins.[25]

It is easy to see that any pair $i, j$ of disks whose separation falls in the range

$$a \leq r_{ij} < 2^{1/2} a \qquad (4.8)$$

are necessarily near neighbors. Figure 4 illustrates such a pair close to the permissible upper distance limit. The two radius-$a$ repulsion circles intersect at cusps that are indicated by arrows in the drawing. The particle positions $\mathbf{r_i}$ and $\mathbf{r_j}$, along with these cusps, form nearly a square. If $r_{ij}$ were to exceed $2^{1/2}a$, the cusps would be closer than this separation. Conceivably a third and a fourth disk could simultaneously take up residence at these cusps, and would have to be paired in preference to $i, j$ pairing. Restricting $r_{ij}$ to the range (4.8) eliminates this possible violation.

Two-particle analogs of $f^{(3)}$ and $g^{(3)}$ are the functions $f^{(2)}(r_{12})$ and $g^{(2)}(r_{12})$. The latter is the conventional pair correlation function,[29] while the former (when multiplied by $r^2 d\mathbf{r_1} d\mathbf{r_2}$) represents the probability that a pair of differential area elements is simultaneously occupied by near neighbors. The implication of the preceding paragraph is that

$$f^{(2)}(r_{12}) = g^{(2)}(r_{12}) \qquad (r_{12} < 2^{1/2} a) . \qquad (4.9)$$

Each near-neighbor pair link serves simultaneously as a side of two contiguous triangles, so for any $r_{12}$ the pair function $f^{(2)}(r_{12})$ can be obtained from $f^{(3)}$ by integration over the superfluous variables:

$$f^{(2)}(r_{12}) = 2r \int_a^\infty dr_{13} \int_a^\infty dr_{23}\ r_{13} r_{23}\ f^{(3)}(r_{12}, r_{13}, r_{23})[M(r_{12}, r_{13}, r_{23})]^{-1/2} \quad . \qquad (4.10)$$

It should be noted in passing that over the range of distances for which both Eqs. (4.7) and (4.9) apply, $f^{(3)}$ contains information to allow evaluation of the ratio

$$K(r,s,t) = g^{(3)}(r,s,t) / g^{(2)}(r) g^{(2)}(s) g^{(2)}(t) \qquad (4.11)$$

for the rigid disk system; this ratio is assigned the value unity by the Kirkwood superposition approximation.[29, 31]

The equation of state of the classical rigid disk system can be expressed in terms of $g^{(2)}(a)$, its contact pair correlation function:[4]

$$bp = r + (p/2) r^2 a^2 g^{(2)}(a) \quad . \qquad (4.12)$$

Here $p$ is the two-dimensional pressure, and $b$ is $(k_B T)^{-1}$. Identity (4.9) above permits this equation to be alternatively written with $f^{(2)}(a)$. Upon inserting the integral contraction relation (4.10) and rearranging, the result connects $f^{(3)}$ directly to the pressure:

$$\frac{1}{pr^2a^2}\left(\frac{bp}{r}-1\right) = \int_a^\infty ds \int_a^\infty dt \, \frac{st\, f^{(3)}(a,s,t)}{[M(a,s,t)]^{1/2}} \quad . \tag{4.13}$$

Under the presumption of a conventional first-order melting/freezing transition, let $r_f$ and $r_c$ respectively be the coexisting fluid and crystal number densities. If the overall number density lies between these values,

$$r_f < r < r_c \quad , \tag{4.14}$$

then the system will display distinct macroscopic regions inhabited by fluid and by crystal, with a negligible amount of material consigned to the interface between these regions. The expected area fractions $a_f$ and $a_c = 1 - a_f$ occupied by these regions are

$$a_f = (r_c - r)/(r_c - r_f) \quad ,$$

$$\tag{4.15}$$

$$a_c = (r - r_f)/(r_c - r_f) \quad .$$

The triangle distribution function in this density interval (4.14) is then equal to a linear combination of $f_f^{(3)}$ and $f_c^{(3)}$, its respective values at $r_f$ and $r_c$:

$$r^3 f^{(3)} = a_f r_f^{\,3} f_f^{(3)} + a_c r_c^{\,3} f_c^{(3)} \quad . \tag{4.16}$$

This result is analogous to those that are known to apply to the correlation functions $g^{(n)}$.[32] Furthermore, the linear combination must obey the virial identity (4.13) across interval (4.14) with a constant pressure.

Finally, we note that $f^{(3)}$ obeys disk confluence conditions that stem from analogous properties of the conventional pair and triplet correlation functions. If $v(r_{ij})$ formally represents the singular rigid-disk pair potential, these latter properties may be expressed as follows:[33-35]

$$\lim_{r_{12} \to 0} \exp[\boldsymbol{b}v(r_{12})] g^{(2)}(r_{12}) = \exp(\boldsymbol{bm}_{ex}) \quad ; \qquad (4.17)$$

$$\lim_{r_{12} \to 0} \exp[\boldsymbol{b}v(r_{12})] g^{(3)}(\mathbf{r}_1, \mathbf{r}_2, \mathbf{r}_3) = \exp(\boldsymbol{bm}_{ex}) g^{(2)}(r_{13}) \quad . \qquad (4.18)$$

Here $\boldsymbol{m}_{ex}$ is the excess (nonideal) chemical potential of the rigid disk system, which can be identified as a reversible work necessary to create a circular cavity large enough to accommodate a disk.[36, 37] These identities (4.17) and (4.18) arise from the obvious fact that two disks constrained to be coincident behave precisely as a single disk so far as the remaining disks are concerned.

Equation (4.9) above indicates that $g^{(2)}$ in Eq. (4.17) may immediately be replaced by $f^{(2)}$:

$$\lim_{r \to 0} \exp[+\boldsymbol{b}v(r)] f^{(2)}(r) = \exp(\boldsymbol{bm}_{ex}) \quad . \qquad (4.19)$$

An analogous reduction for Eq. (4.18) requires that triplet disk configurations used for the limit operation belong to the set **U**. If this is so, then we have

$$\lim_{r \to 0} \exp[bv(r)] f^{(3)}(r,s,t) \equiv \lim_{r \to 0} \exp[bv(r)] f^{(3)}(r,s,s)$$
$$= \exp(bm_{ex}) f^{(2)}(s) \quad (4.20)$$

V. Approach to Close Packing

Rigid disks in the plane attain their maximum packing density in a regular triangular lattice, with each disk touching six neighbors.[38] In this arrangement

$$ra^2 \equiv r_0 a^2$$
$$= 2/3^{1/2} \quad (5.1)$$
$$= 1.1547.... \quad .$$

Computer simulation studies[8] suggest that when $ra^2$ lies between approximately 0.91 and this upper limit, the thermodynamically stable phase of large disk systems is the corresponding triangular crystal with increased mean neighbor spacing that allows restricted local motion. The equilibrium concentration of defects, specifically vacancies, appears to be very small throughout this density range, and to vanish exponentially as $r$ approaches $r_0$.[39]

It is generally accepted [40, 41] that the disk crystal pressure exhibits a simple-pole divergence at close packing:

$$bpa^2 = s/(1 - r/r_0) + O[(1 - r/r_0)^0] ,$$
$$s = 4/3^{1/2} = 2.3094.... \quad . \quad (5.2)$$

The exact form of $f^{(3)}$ must be consistent with this behavior, through identity (4.13). At the same time it must continue to conform to the other two basic identities, Eqs. (4.2) and (4.5).

The Delaunay triangulation of the highly compressed disk crystal consists everywhere of nearly equilateral triangles. Their side lengths are narrowly distributed just beyond the collision distance $a$. The difference $l$ between the mean crystal spacing, and $a$, provides a natural length scale for this narrow distribution:

$$l(\mathbf{r}) = a[(\mathbf{r}/\mathbf{r}_0)^{1/2} - 1] \quad . \tag{5.3}$$

Comparison with the previous Eq. (5.2) shows that in the high compression limit,

$$l(\mathbf{r})/a \sim \mathbf{s}/2\mathbf{b}pa^2 \quad . \tag{5.4}$$

The triangle distribution function $f^{(3)}$ must adopt a simple scaled form in this limit, which we choose to write in the following way:

$$f^{(3)}(r,s,t) \sim (a/l)^3 \, \mathbf{f}[(r-a)/l, (s-a)/l, (t-a)/l] \quad . \tag{5.5}$$

The symmetric function $\mathbf{f}(u,v,w)$ should be independent of $a$ and of density, as written.

When expression (5.5) is inserted into either of the basic identities (4.2) or (4.5), with appropriate limiting forms for the other integrand factors, the result is

$$\frac{3^{5/2}}{8\mathbf{p}} = \int_0^\infty du \int_0^\infty dv \int_0^\infty dw \, \mathbf{f}(u,v,w) \quad . \tag{5.6}$$

Treated similarly, the pressure identity (4.13) yields a different integral condition that must be satisfied by $\mathbf{f}$:

$$\frac{9\mathbf{s}}{16\mathbf{p}} = \int_0^\infty dv \int_0^\infty dw\, \mathbf{f}(0,v,w) \quad . \tag{5.7}$$

Asymptotic result (5.4) has been used to eliminate pressure from this last equation.

VI. Minimal Closure Approximation

The identities contained in Section IV above convey basic information upon which a self-consistent $f^{(3)}$ calculation can be crafted. We now examine a low-order, essentially minimal, closure approximation for purposes of illustration. Determining whatever shortcomings it may possess should assist in a subsequent search for a more accurate theory of rigid-disk short-range order and equation of state.

Our closure approximation choice emphasizes the fundamental role played by $R$, the circumscribed circle radius. It postulates the following form for $f^{(3)}$:

$$f^{(3)} \cong C\exp[-\mathbf{p}\mathbf{b}pR^2(r,s,t) - 2\mathbf{p}\mathbf{b}\mathbf{g}R(r,s,t)] \quad . \tag{6.1}$$

Here $\mathbf{g}$ represents a line tension for the rigid disk system at an impenetrable linear boundary. Expression (6.1) has a form suggested by the two-dimensional scaled particle theory,[37] *i.e.* a suitably normalized Boltzmann factor for the reversible work (against pressure and boundary tension) to empty a circular cavity of radius $R$. Note that any $O(R^0)$ contribution to this work would automatically be absorbed into the normalization constant $C$.

Approximation (6.1) becomes exact in the low density limit. The Appendix demonstrates that this ideal gas limit has form (6.1) with $\boldsymbol{g} = 0$ and $C = 1$. The effects of disk interactions at positive density enter through $\boldsymbol{g}$ as well as through the requirement that triangle side lengths must equal or exceed $a$.

Expression (6.1) contains three scalar unknowns, $C$, $p$, and $\boldsymbol{g}$. Consequently three independent conditions are required for self-consistent solution as a function of density. We have chosen the basic identities (4.2), (4.5), and (4.13) to fulfill this role. The resulting nonlinear mathematical problem and its numerical analysis are described in the following Section VII.

Before leaving this Section VI, we wish to emphasize that closure approximation (6.1) has the capacity to describe the approach to the close-packed limit, as well as the ideal gas limit. As discussed earlier in Section V, high compression forces all triangle sides to remain close to contact distance $a$. Upon making the appropriate expansion of $R$, Eq. (2.1), in these small increments, it is possible to show that the $f^{(3)}$ scaling function $\boldsymbol{f}$ introduced in Eq. (5.5) is the following:

$$\boldsymbol{f}(u,v,w) = C_f \exp\{-(2\boldsymbol{p}\boldsymbol{b}l/9a)(pa^2 + 3^{1/2}\boldsymbol{g}a)(u+v+w)\} ,$$
$$C_f = 3^{-7/2}\boldsymbol{p}^2[\boldsymbol{b}(pa^2 + 3^{1/2}\boldsymbol{g}a)]^3(l/a)^3 .$$
(6.2)

The value shown for the normalization constant $C_f$ assures that Eq. (5.6) will be satisfied. Equation (5.7) then requires

$$\bm{b}(pa^2 + 3^{1/2}\bm{g}a) \sim (3^{3/2}\bm{s}/4\bm{p})(a/l) \ . \tag{6.3}$$

VII. Numerical Solution

Disk collision diameter $a$ is the natural length unit choice. With this convention, topological identity (4.2) can be used to evaluate normalization constant $C(\bm{r})$ in expression (6.1), for any given values of $\bm{b}p$ and $\bm{b}g$:

$$C(\bm{r}) = 3/\{2\bm{p}\bm{r}^2 \int_1^\infty dr \int_1^\infty ds \int_1^\infty dt \ rst \ [M(r,s,t)]^{-1/2} \\ \times \exp[-\bm{p}\bm{b}pR^2(r,s,t) - 2\bm{p}\bm{b}gR(r,s,t)]\} \ . \tag{7.1}$$

This reduces the numerical problem to a search over the half plane

$$\bm{b}p > 0 \ , \qquad -\infty < \bm{b}g < +\infty \ , \tag{7.2}$$

for solutions $\bm{b}p$, $\bm{b}g$ as functions of $\bm{r}$ that satisfy the two remaining identities of choice, Eqs. (4.5) and (4.13). In the same units convention, these become the following pair of coupled nonlinear integral equations:

$$3/\bm{p}\bm{r}^3 = C \int_1^\infty dr \int_1^\infty ds \int_1^\infty dt \ rst \exp[-\bm{p}\bm{b}pR^2(r,s,t) - 2\bm{p}\bm{b}gR(r,s,t)] \ ; \tag{7.3}$$

$$[(\bm{b}p/\bm{r}) - 1]/\bm{p}\bm{r}^2 = C \int_1^\infty ds \int_1^\infty dt \ st \ [M(1,s,t)]^{-1/2} \\ \times \exp[-\bm{p}\bm{b}pR^2(1,s,t) - 2\bm{p}\bm{b}gR(1,s,t)] \ . \tag{7.4}$$

Numerical solutions to the last two equations have been sought using a commercial software package.[42] We have found that a single solution, continuous in $r$, exists over the entire density range $0 < r < r_0$. Computed results appear graphically in Figures 5 and 6.

The reduced pressure *vs.* reduced density is shown in Figure 5 as a solid curve. The scaled particle theory result,[37]

$$bpa^2 = ra^2(1 - pra^2/4)^{-2} \qquad \text{(SPT)}, \qquad (7.5)$$

shown as a dashed curve, has been included for comparison. While both theories at low density are correct through the third virial coefficient, neither predicts a phase transition. Although scaled particle theory provides an excellent account of the fluid-phase pressure,[43] its divergence occurs at a density that exceeds the close-packed limit; furthermore its singularity is a double pole, not a simple pole as required by Eq. (5.2). By contrast, the present approach predicts fluid-range pressures that are somewhat low, while a simple-pole divergence occurs properly at $r_0$ (though its residue numerically is approximately 3.4 instead of the correct value 2.3….).

We have also included in Figure 5 an indication of the freezing transition location reported by Alder and Wainwright,[6] who have inferred that a simple first-order phase change is involved. In a rough sense, the horizontal coexistence region for this transition seems to connect the two theoretical curves.

Figure 6 shows the boundary tension predictions for the present approach (solid curve) and from the scaled-particle theory (dashed curve):[37]

$$b g a = - \frac{2}{p} \left( \frac{p r a^2 / 4}{1 - p r a^2 / 4} \right)^2 \qquad \text{(SPT)} . \qquad (7.6)$$

Although both are negative and vanish at zero density, the discrepancy is otherwise substantial. As in the case of pressure, the scaled particle result remains finite at close packing, while the present theory produces a simple-pole divergence there. No direct, independent determination of the rigid disk $g$ is available to assess the relative merits of these two predictions over the intermediate density range.

The Kirkwood superposition factor for mutual contact, $K(a,a,a)$, adopts a simple form when closure approximation (6.1) is accepted:

$$K(a,a,a) = C(p r a^2 / 2)^3 [(b p / r) - 1]^{-3} \\ \times \exp(-p b p a^2 / 3 - 2 p b g a / 3^{1/2}) \quad . \qquad (7.7)$$

Figure 7 shows that this quantity is close to, but somewhat greater than, unity over the full density range. We are not aware of any other determinations of this factor that would provide a direct comparison. However it might be noted that molecular dynamics simulation results for rigid spheres in three dimensions indicate that the same three-particle quantity $K(a,a,a)$ indeed remains close to unity at all densities, including those in the crystalline region.[44]

VIII. Conclusions and Discussion

The most obvious shortcoming of closure approximation Eq. (6.1) is its failure to produce a freezing transition of any thermodynamic order. Nevertheless it can be valuable if it helps to identify a key missing ingredient or concept that ought to be incorporated in a successor theory. By hindsight, we conclude that the circumscribed circle radius $R$ does not alone provide a sufficiently complete classification of the near-neighbor triangles in the rigid disk system.

Although $R$ is a convenient measure of triangle size, it is insensitive to triangle shape. Provided $R > a/3^{1/2}$, the three disks can be rearranged on the circumscribed circle into a variety of shapes of different symmetries. The relevance of shape emerged in a recent molecular dynamics study of textural patterns in the disk system as the fluid phase was compressed toward the freezing point:[19] nearly equilateral triangles cluster into islands embedded in a sea of more irregular triangles. Freezing can be described as a process that eliminates the latter.

The following dimensionless functions of triangle side lengths measure shape irregularity:

$$L_1(r,s,t) = \frac{(r-s)^2 + (r-t)^2 + (s-t)^2}{r^2 + s^2 + t^2} \quad ; \quad (8.1)$$

$$L_2(r,s,t) = \frac{(r-s)^2(r-t)^2(s-t)^2}{(r^2 + s^2 + t^2)^3} \quad . \quad (8.2)$$

Both are non-negative and vanish for equilateral triangles. The first is positive for all other shapes, while the second remains zero for isosceles triangles. An attractive option for future study would be to include an additional term of type $lL(r,s,t)$ in the exponent of closure approximation (6.1) to build in shape sensitivity, at the cost of another scalar unknown ($l$) and another determining condition from Section IV.

Assuming that an $f^{(3)}$ approximation can be found that produces at least a satisfactory fluid phase description, it is worth mentioning that the present approach is applicable to mixtures. This extension would require consideration of a set of triangle distribution functions $f_{\alpha\beta\gamma}^{(3)}$, indexed by the species $\alpha$, $\beta$, $\gamma$ of the vertices. Attention need not be restricted to the case of additive diameters, and in the binary mixture with

$$a_{ab} > (a_{aa} + a_{bb})/2 \qquad (8.3)$$

it should be feasible to search for the presence of immiscible phases.

Extension of the present approach to rigid spheres in three dimensions presents a significant challenge. First, the basic units of the Voronoi-Delaunay tiling of space are tetrahedra, so the distribution function of interest would be $f^{(4)}$, a function of six tetrahedron edge lengths. Second, the near-close-packed crystals (FCC, HCP, and their hybrids) contain two distinct types of nondegenerate (positive volume) tetrahedra: one has all six edges near contact, the other has one of its edges $2^{1/2}$ times as long as the remaining five edges. Furthermore, these two tetrahedron types must be present in exactly 1:2 ratio respectively in order to form a close-packed crystal.

Finally, we stress that a comprehensive understanding of the $f^{(3)}$ description for the rigid disk system (more generally the $f^{(d+1)}$ description for rigid "hyperspheres" in $d$ dimensions) requires analysis of the Helmholtz free energy functional $F_N\{f^{(3)}\}$ (respectively $F_N\{f^{(d+1)}\}$). For given density and temperature this must be at an absolute minimum with respect to all permissible variations of $f^{(3)}$. The corresponding functional for rigid rods in one dimension has an elementary exact form:

$$\beta F_N\{f^{(2)}\} = N[\ln(\lambda_T \rho) + \rho \int_a^\infty f^{(2)}(x) \ln f^{(2)}(x)\, dx] \quad, \qquad (8.4)$$

where $f^{(2)}(x)$ is the neighbor pair separation distribution, and $l_T$ is the mean thermal de Broglie wavelength; all $f^{(2)}$ variations are permissible subject only to normalization and mean separation conditions. However the situation is a bit more subtle for $d = 2$ on account of the basic issue of precisely what are the geometrically allowable distribution functions $f^{(3)}$. It is easy to find functions $f^{(3)}$ that are inconsistent with the near-neighbor disk pairing procedure, and thus are not allowable. Assuming that this aspect of the rigid disk problem can be mastered, one should be in a better position to judge the appropriateness of candidate $f^{(3)}$ approximations to be used in subsequent self-consistent theories of rigid disk short-range order and equation of state.

Appendix

Setting collision diameter $a$ to zero reduces the disk system to a classical two-dimensional ideal gas. So far as the triangle distribution function $f^{(3)}$ is concerned, only the circumscribed circle is relevant. As described in Section II above, this circle is an inviolable zone that must not be penetrated by any of the $N-3$ particles not involved in the near-neighbor triangle under consideration. But otherwise these $N-3$ particles are unconstrained.

A general result in classical equilibrium statistical mechanics is that the probability for some region to be devoid of particles as a result of a local density fluctuation is proportional to $\exp(-\beta W)$, where $\beta = (k_B T)^{-1}$ and $W$ is the reversible isothermal work that would have to be expended to empty that region.[45] In the case of a circular region with radius $R$, embedded in the two-dimensional ideal gas,

$$W = p \cdot \mathbf{p} R^2 \quad , \tag{A.1}$$

the pressure-area product. For this ideal gas of course we have the equation of state:

$$\mathbf{b} p = \mathbf{r} \quad . \tag{A.2}$$

These considerations require that $f^{(3)}$ for the two-dimensional ideal gas have the form:

$$f^{(3)}(r,s,t) = \exp[-\mathbf{pr} R^2(r,s,t)] \quad , \tag{A.3}$$

where Eq. (2.1) above gives the explicit expression for $R$ in terms of the three triangle sides. Notice that the pre-exponential factor in the right member of Eq. (A.3) is unity, since any three particles with vanishing $R$ must be mutual near neighbors with unit probability.

Verifying that expression (A.3) obeys the basic identities (4.2), (4.5), and (4.13) proceeds most directly by introducing a variable change. This is motivated by the triangle inequalities

$$\begin{aligned} r &< s+t \ , \\ s &< r+t \ , \\ t &< r+s \ , \end{aligned} \tag{A.4}$$

that limit the range of these side-length variables to the interior of a three-fold-symmetric pyramidal region in the first octant, as illustrated by Figure 8(a). A simple change of variables,

$$x = r^2 \ ,$$
$$y = s^2 \ ,  \quad\quad\quad\quad (A.5)$$
$$z = t^2 \ ,$$

converts that pyramidal region to a circular cone whose sides are tangent to the coordinate planes, as shown in Figure 8(b). The cone's vertex angle $\theta_m$ is easily found to be

$$\cos\theta_m = (2/3)^{1/2} \ , \quad\quad\quad\quad (A.6)$$
$$\theta_m \cong 35.27° \ .$$

With this simpler boundary for the geometrically allowed region, a subsequent transformation to spherical polar coordinates $r_0, \theta, \varphi$ measured about the cone axis becomes natural.

In terms of the last variable set,

$$f^{(3)} = \exp\left\{-\left(\frac{pr_0}{6\cdot 3^{1/2}}\right)\left(\frac{5\cos^3\theta - 3\cos\theta + 2^{1/2}\sin^3\theta \sin(3\varphi)}{3\cos^2\theta - 2}\right)\right\} \ , \quad\quad\quad\quad (A.7)$$

and

$$M = r_0^2(3\cos^2\theta - 2) \ . \quad\quad\quad\quad (A.8)$$

Proceeding to identity (4.2), the integral forming its right member now appears as follows:

$$\int_0^\infty dr \int_0^\infty ds \int_0^\infty dt \ rst \ f^{(3)} \ M^{-1/2}$$
$$= \frac{1}{8} \int_0^\infty r_0 \, dr_0 \int_0^{q_m} \sin q \, dq \int_0^{2p} dj \ f^{(3)}(r_0, q, j)(3\cos^2 q - 2)^{-1/2} \ . \quad (A.9)$$

The $r_0$ integration is elementary and can be carried out immediately to yield

$$\frac{27}{2p^2 r^2} \int_0^{q_m} \sin q \, dq \int_0^{2p} dj \ \frac{(3\cos^2 q - 2)^{3/2}}{[5\cos^3 q - 3\cos q + 2^{1/2} \sin^3 q \sin(3j)]^2} \ . \quad (A.10)$$

The $j$ integral is a standard form,[46] so that expression (A.10) leads to

$$\frac{27}{pr^2} \int_0^{q_m} \sin q \, dq \ \frac{(5\cos^3 q - 3\cos q)}{(3\cos^2 q - 1)^3}$$
$$= \frac{27}{2pr} \int_{2/3}^1 dx \ \frac{5x - 3}{(3x - 1)^3} \quad (A.11)$$
$$= \frac{3}{2pr^2} \ ,$$

precisely the result required to satisfy identity (4.2).

The same variable transformation strategy is applicable to the mean-area identity (4.5). Its integral right member, for the ideal gas in two dimensions, undergoes the simplification:

$$\int_0^\infty dr \int_0^\infty ds \int_0^\infty dt \ rst \ f^{(3)} = (324 \cdot 3^{1/2} / p^2 r^3) I \ , \quad (A.12)$$

where

$$I = \int_{(2/3)^{1/2}}^{1} du \ (3u^2 - 2)^{1/2} (3u^2 - 1)^{-5} (24u^6 - 27u^4 + 6u^2 + 1) \quad . \tag{A.13}$$

This remaining single integral is not elementary, but because the exact $f^{(3)}$, Eq. (A.3), necessarily must obey identity (4.5) we know that

$$\begin{aligned} I &= p/108 \cdot 3^{1/2} \\ &\cong 1.67944 \times 10^{-2} \quad , \end{aligned} \tag{A.14}$$

and indeed accurate numerical evaluation of $I$ serving as a cross check confirms that it is so.

The virial pressure identity (4.13) becomes formally indeterminate in the ideal gas limit, and so will not be considered in that form. However the correct limiting behavior can effectively be extracted from the closely related expression (4.10) for the near-neighbor probability function $f^{(2)}$. Upon using transformation (A.5) to effect the conical geometry of Figure 8(b), we have

$$\begin{aligned} f^{(2)}(x^{1/2}) = \frac{r}{2} \int dy \int dz \ &\exp[-prxyz/M(x^{1/2}, y^{1/2}, z^{1/2})] \\ &\times [M(x^{1/2}, y^{1/2}, z^{1/2})]^{-1/2} \quad , \end{aligned} \tag{A.15}$$

where the integration spans the region over which $x$, $y$, and $M$ are simultaneously positive. The restriction to fixed $r$ (i.e. fixed $x$) in this integral corresponds to the intersection of the cone shown in Figure 8(b) with a constant-$x$ plane that is parallel to its side; the resulting conical section is a parabola. It now becomes natural to use $M$ itself as an integration variable, along with

$$X = 2^{-1/2}(y - z) \quad , \tag{A.16}$$

for which the transformation Jacobian is

$$\frac{\partial(M,X)}{\partial(y,z)} = \det\begin{pmatrix} 2(x-y+z) & 2(x+y-z) \\ 2^{-1/2} & -2^{-1/2} \end{pmatrix} \qquad (A.17)$$
$$= 2^{3/2} x \quad .$$

This permits Eq. (A.15) to undergo substantial simplification:

$$f^{(2)}(x^{1/2}) = (\boldsymbol{r}/2^{5/2}x)\exp(-\boldsymbol{p}\boldsymbol{r}x/8)\int_{-\infty}^{\infty} dX\ \exp(-\boldsymbol{p}\boldsymbol{r}X^2/4x)$$
$$\times \int_0^{\infty} dM\ M^{-1/2}\exp\{-(\boldsymbol{p}\boldsymbol{r}/16x)[M+(2X^2-x^2)^2/M]\} \quad . \qquad (A.18)$$

The *M* integral is a standard form and may be carried out explicitly, followed by the same for the *X* integral. The result is

$$f^{(2)}(r) = \boldsymbol{r}^{1/2}r\exp(-\boldsymbol{p}\boldsymbol{r}r^2/4) + erfc[(\boldsymbol{p}\boldsymbol{r})^{1/2}r/2] \quad , \qquad (A.19)$$

where

$$erfc(z) = (2/\boldsymbol{p}^{1/2})\int_z^{\infty} du\ \exp(-u^2) \qquad (A.20)$$

is the error function complement. The result (A.19) was derived previously by Collins via a different route.[25]

As expected, expression (A.19) verifies that $f^{(2)}$ is unity at zero separation, and declines monotonically to zero with increasing $r$. It can be used to calculate moments of the neighbor separation $r$, for example

$$\begin{aligned}\langle r \rangle &= 32/9 \boldsymbol{pr}^{1/2} \\ &\cong 1.1318/\boldsymbol{r}^{1/2} \quad .\end{aligned} \qquad (A.21)$$

This can be compared to the corresponding result when all particles form a perfect triangular lattice, for which

$$\begin{aligned}\langle r \rangle &= 2^{1/2}/3^{1/4}\, \boldsymbol{r}^{1/2} \\ &\cong 1.0746/\boldsymbol{r}^{1/2} \quad .\end{aligned} \qquad (A.22)$$

Acknowledgements

The authors are delighted to contribute this paper to the "Festschrift" recognizing George Stell's seminal contributions to liquid-state theory. S.T. is indebted to George as a mentor, and for the love of research that he imparted to him.

S.T. gratefully acknowledges the support of the U.S. Department of Energy, Office of Basic Energy Sciences (Grant No. DE-FG02-92ER14275). T.M.T. acknowledges the National Science Foundation for financial support. P.G.D. gratefully acknowledges support of the U.S. Department of Energy, Office of Basic Energy Sciences (Grant No. DE-FG02-87ER13714), and the donors of the Petroleum Research Fund, administered by the American Chemical Society.

References

1. L. Boltzmann, *Verslag. Gewone Vergadering Afd. Natuurk. Nederlandse Akad. Wetensch.* **7**: 484 (1899).

2. B. R. A. Nijboer and L. Van Hove, *Phys. Rev.* **85**: 777 (1952).

3. J. D. Bernal, *Proc. Roy. Inst. Gt. Brit.* **37** (*No. 168*): 355 (1959); also *Proc. Roy. Soc. Ser A* **280**: 299 (1964).

4. H.L. Frisch, *Adv. Chem. Phys.* **VI**: 229 (1964).

5. J.O. Hirschfelder, C.F. Curtiss, and R.B. Bird, *Molecular Theory of Gases and Liquids* (Wiley, New York, 1954).

6. B.J. Alder and T.E. Wainwright, *Phys. Rev.* **127**: 359 (1962).

7. W.W. Wood and J.D. Jacobson, *J. Chem. Phys.* **27**: 1207 (1957).

8. (a) A.C. Mitus, H. Weber, and D. Marx, *Phys. Rev. E* **55**: 6855 (1997); (b) A. Jaster, *Europhys. Lett.* **42**: 277 (1998).

9. G. Mason, *Disc. Faraday Soc.* **43**: 75 (1967).

10. J.L. Finney, *Mater. Sci. Eng.* **23**: 199 (1976).

11. E.L. Hinrichsen, J.Feder, and T. Jossang, *Phys. Rev. A* **41**: 4199 (1990).

12. (a) B.D. Lubachevsky, F.H. Stillinger, and E.N. Pinson, *J. Stat. Phys.* **64**: 501 (1991); (b) R.J. Speedy, *J. Phys. Condensed Matter* **10**: 4185 (1998).

13. S. Torquato, *Phys. Rev. Letters* **74**: 2156 (1995).

14. I.C. Sanchez, *J. Chem. Phys.* **101**: 7003 (1994).

15. N. Metropolis, A.W. Rosenbluth, M.N. Rosenbluth, and A.H. Teller, *J. Chem. Phys.* **21**:1087 (1953).

16. J.A. Zollweg and G.V. Chester, *Phys. Rev. B* **46**: 11187 (1992).

17. J. Lee and K.J. Strandburg, *Phys. Rev. B* **46**: 11190 (1992).

18. J.J. Erpenbeck and M. Luban, *Phys. Rev. A* **32**: 2920 (1985).

19. T.M. Truskett, S. Torquato, S. Sastry, P.G. Debenedetti, and F.H. Stillinger, *Phys. Rev. E* **58**: 3083 (1998).

20. J.M. Kosterlitz and D.J. Thouless, *J. Phys. C* **6**: 1181 (1973).

21. B.I. Halperin and D.R. Nelson, *Phys. Rev. Letters* **41**: 121 (1978).

22. D.R. Nelson and B.I. Halperin, *Phys. Rev. B* **19**: 2457 (1979).

23. A.P. Young, *Phys. Rev. B* **19**: 1855 (1979).

24. K.J. Strandburg, *Rev. Mod. Phys.* **60**: 161 (1988).

25. R. Collins, *J. Phys. C* **1**: 1461 (1968).

26. R. Collins, in *Phase Transitions and Critical Phenomena, Vol. 2*, C. Domb and M.S. Green, eds. (Academic Press, New York, 1972), pp. 271-303.

27. M. Tanemura, T. Ogawa, and N. Ogita, *J. Comput. Phys.* **51**: 191 (1983), and background references cited therein.

28. S. Sastry, D.S. Corti. P.G. Debenedetti, and F.H. Stillinger, *Phys. Rev. E* **56**: 5524 (1997).

29. T.L. Hill, *Statistical Mechanics* (McGraw-Hill, New York, 1956), chap. 6.

30. H.S.M. Coxeter, *Regular Polytopes, 2nd edition* (Macmillan, New York, 1963), p.9.

31. J.G. Kirkwood and E.M. Boggs, *J. Chem. Phys.* **10**: 394 (1942).

32. J.E. Mayer and E. Montroll, *J. Chem. Phys.* **9**: 2 (1941).

33. W.G. Hoover and J.C. Poirier, *J. Chem. Phys.* **37**: 1041 (1962).

34. E. Meeron and A.J.F. Siegert, *J. Chem. Phys.* **48**: 3139 (1968).

35. D. Henderson and E.W. Grundke, *Mol. Phys.* **24**: 669 (1972).

36. H. Reiss, H.L. Frisch, and J.L. Lebowitz, *J. Chem. Phys.* **31**: 369 (1959).

37. E. Helfand, H.L. Frisch, and J.L. Lebowitz, *J. Chem. Phys.* **34**: 1037 (1961).

38. C.A. Rogers, *Packing and Covering* (Cambridge Univ. Press, Cambridge, 1964), chap.1.

39. F.H. Stillinger, Z.W. Salsburg, and R.L. Kornegay, *J. Chem. Phys.* **43**: 932 (1965).

40. B.J. Alder, W.G. Hoover, and D.A. Young, *J. Chem. Phys.* **49**: 3688 (1968).

41. F.H. Stillinger and Z.W. Salsburg, *J. Stat. Phys.* **1**: 179 (1969).

42. MATHCAD version 6.0, Mathsoft Corp., Cambridge, MA. Sensitivity of results to integration upper limits, and to solution precision limits, was carefully explored.

43. Compare, for example, the SPT predictions with the accurate fluid equation of state appearing as Eq.(11) in ref. 14.

44. B.J. Alder, *Phys. Rev. Letters* **12**: 317 (1964).

45. H. Reiss, *Adv. Chem. Phys.* **IX**: 1 (1965).

46. I.S. Gradshteyn and I.M. Ryzhik, *Tables of Integrals, Series, and Products* (Academic Press, New York, 1980), p. 383, formula 3.661.4.

Figure Captions

1. Elements of the Voronoi-Delaunay tessellation for particles (black dots) in the plane. The Voronoi near-neighbor polygons are indicated with solid lines, while all pairs of near neighbors are connected by dashed lines.

2. Near-neighbor triangle for particles *i, j,* and *k*, and its circumscribed circle. In order to maintain validity of the three pairings, all other particle positions must be exterior to the circle ($l$), not interior to the circle ($l'$).

3. Extremal configuration of three disks for which the circumscribed circle is just covered by disk exclusion envelopes (radius-*a* circles). The near-neighbor separations *r, s,* and *t* are all equal to $3^{1/2}a$.

4. Limiting pair distance ($r_{ij} = 2^{1/2} a$) beyond which near-neighbor pairing could be interrupted by particles at the cusps (arrows).

5. Pressure equation of state calculated from simultaneous Eqs. (7.1), (7.3), and (7.4) (solid curve). For comparison, the scaled-particle-theory pressure for disks from ref. 37 has been included (dashed curve), as well as the location of the Alder-Wainwright freezing transition from ref. 6.

6. Reduced boundary tension *vs*. reduced density calculated from simultaneous Eqs. (7.1), (7.3), and (7.4) (solid curve). The corresponding scaled-particle-theory result from ref. 37 has been included (dashed curve).

7. Kirkwood superposition factor, Eq. (4.11), for three disks in mutual contact.

8. Integration regions (a) before, and (b) after applying transformation (A.5).

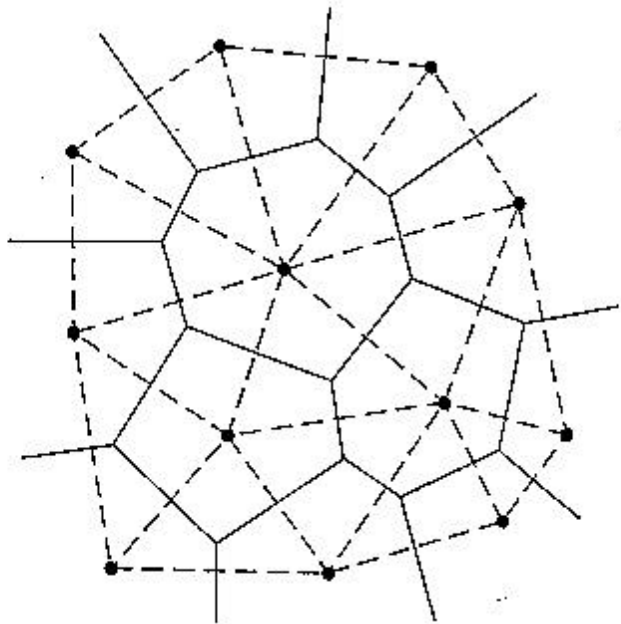

CAD 12053.1

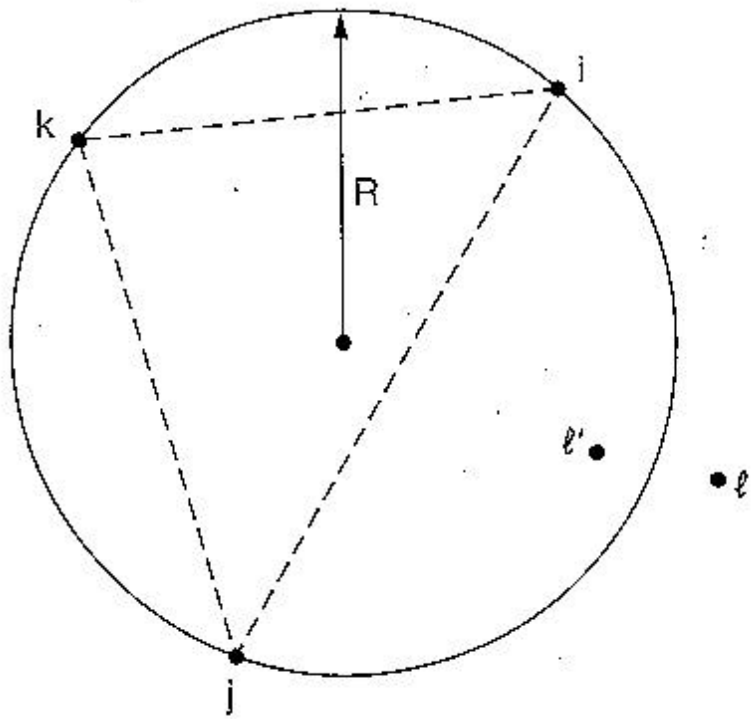

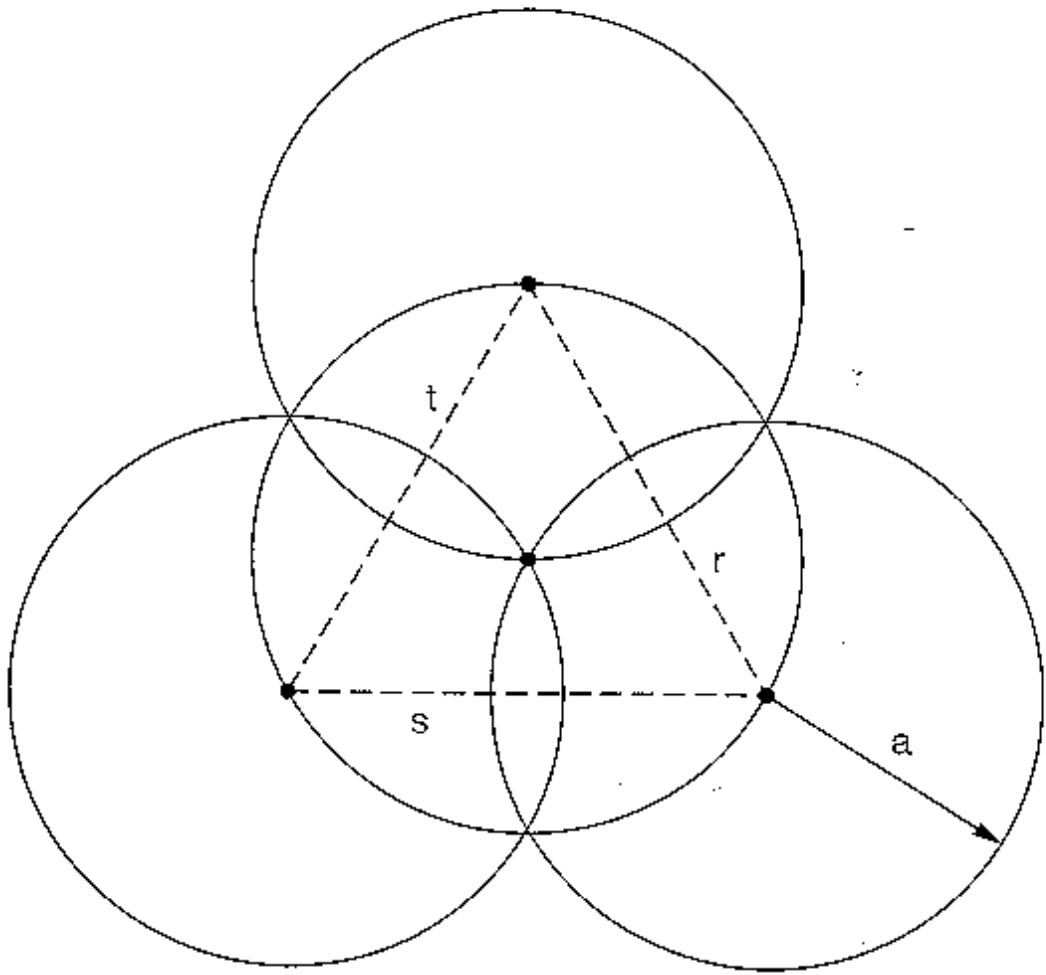

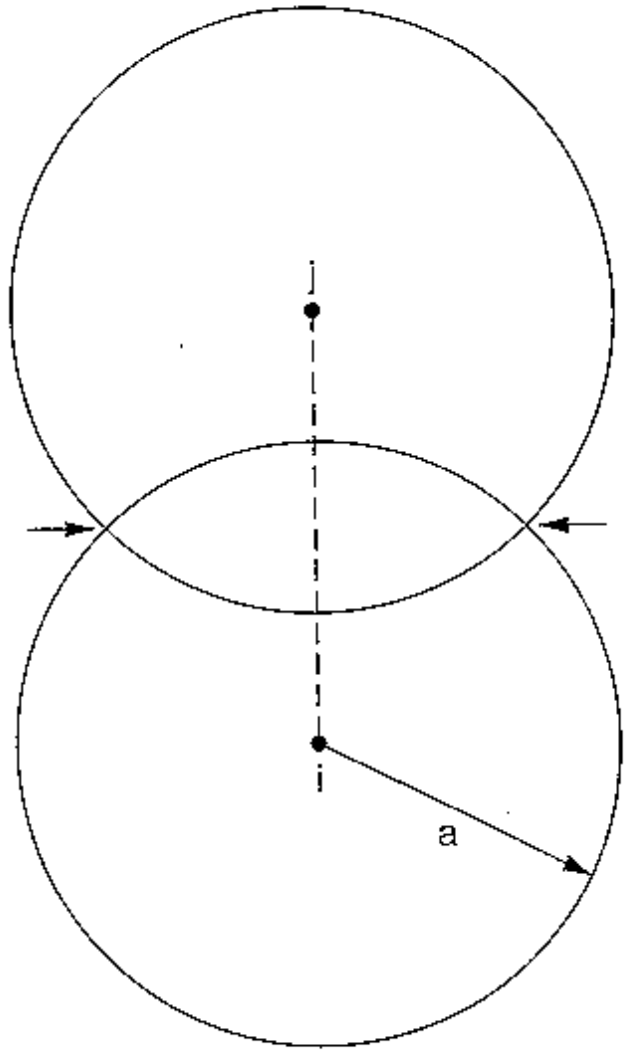

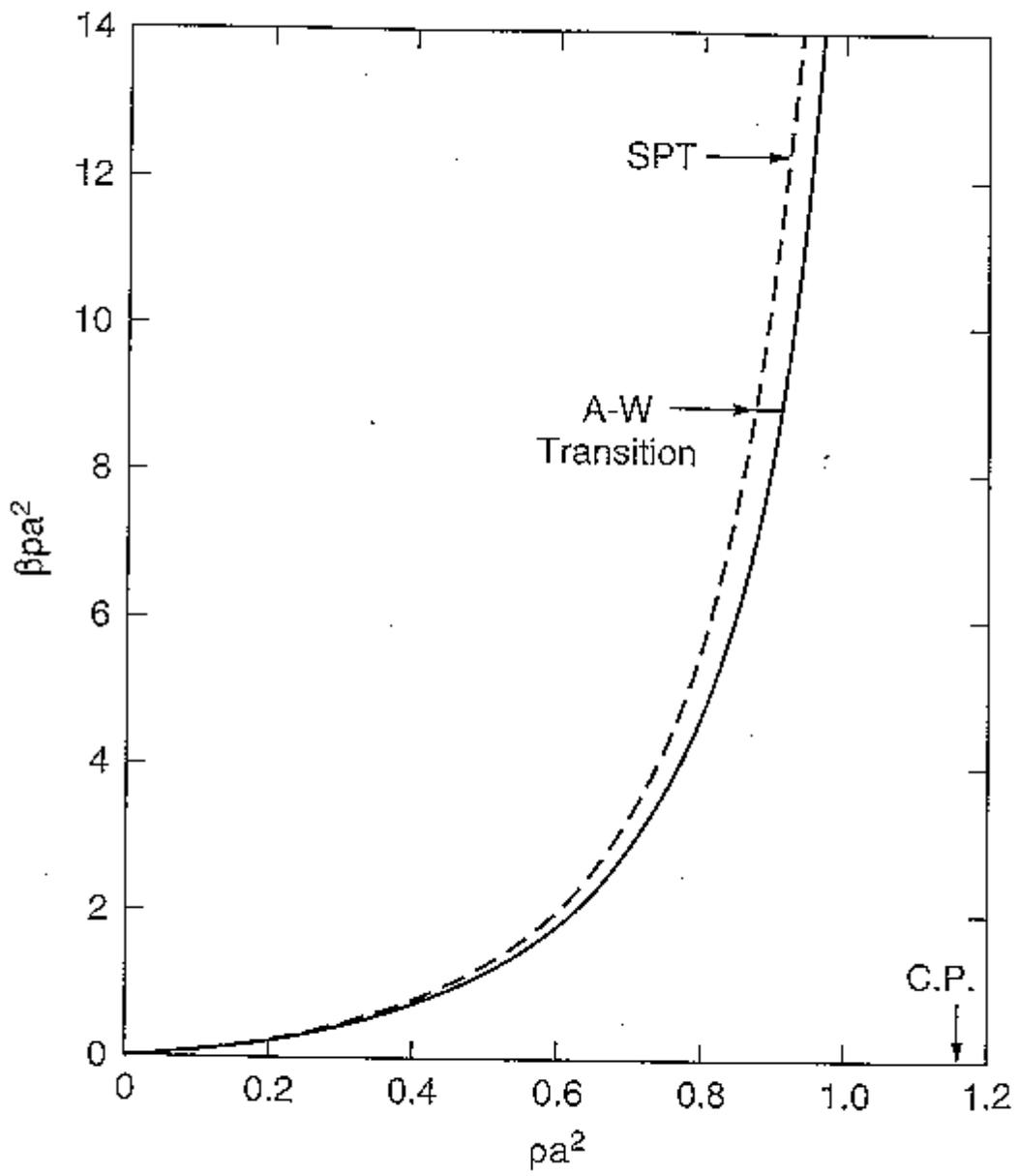

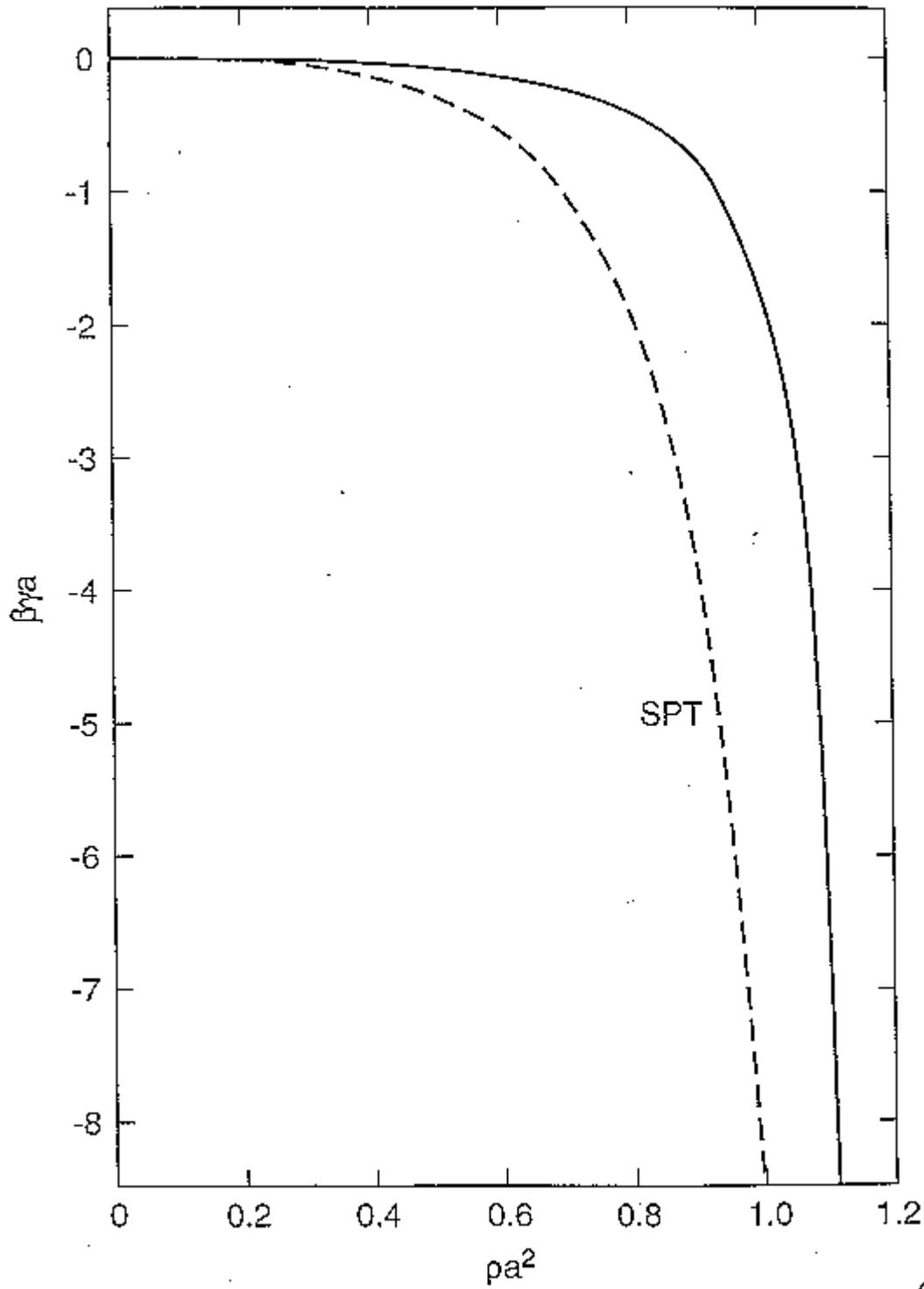

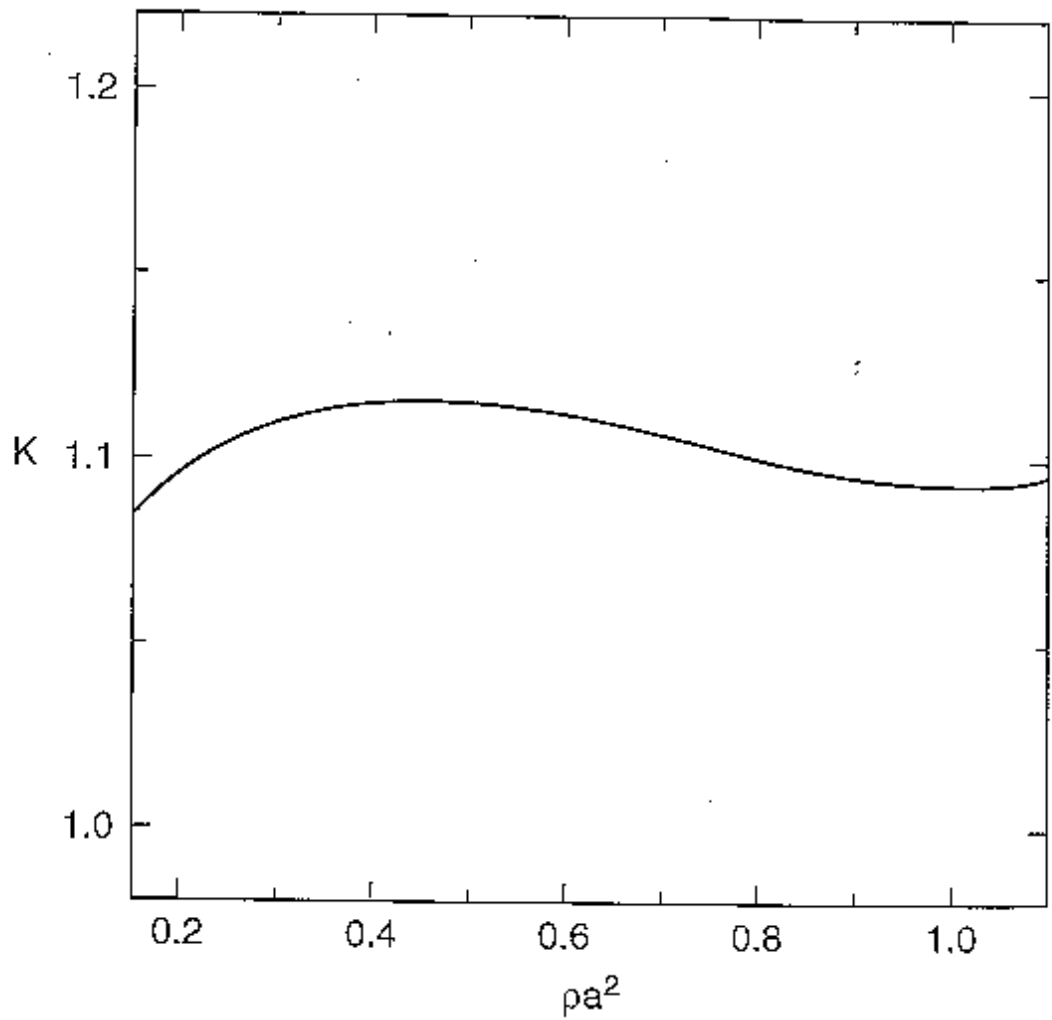

(a)
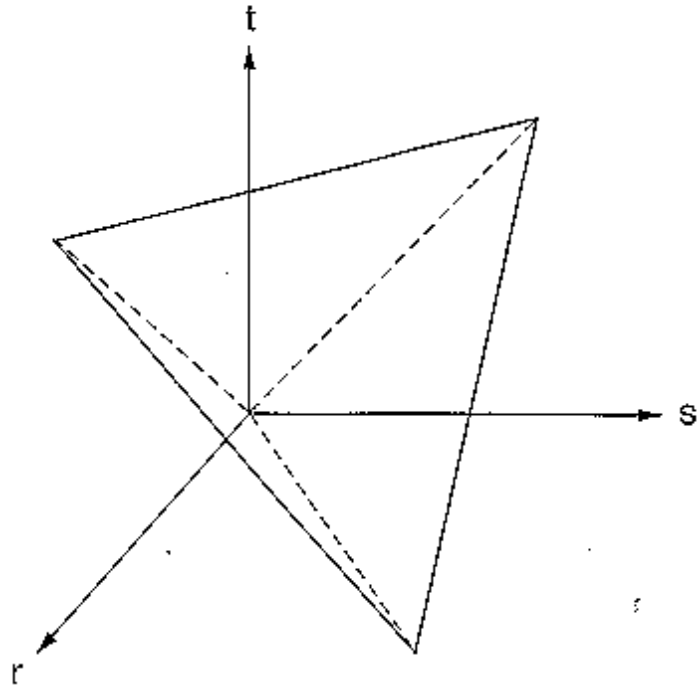

(b)
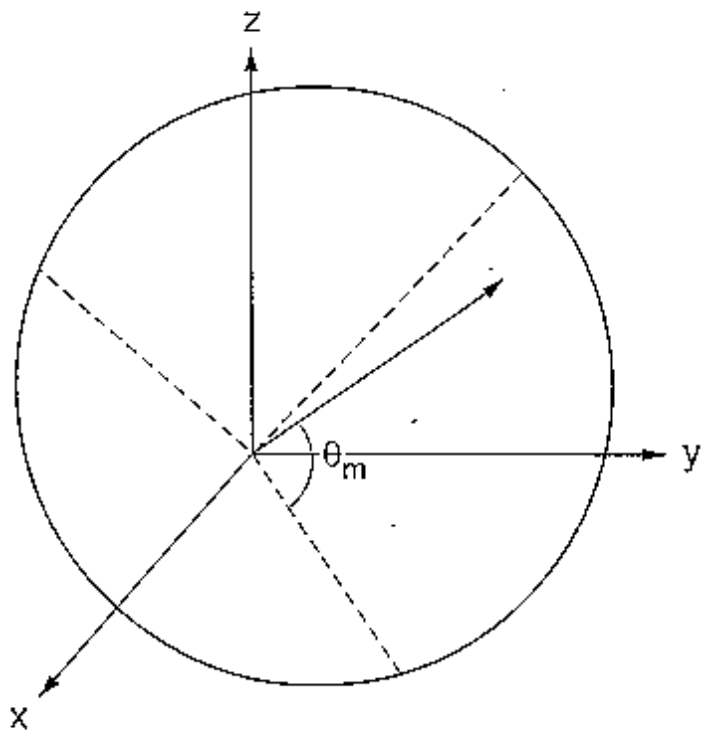